\documentclass[12pt, letterpaper]{JHEP3} 
\usepackage{epsf}
\usepackage{graphicx,epsfig}
\usepackage{amsmath}
\usepackage{amsfonts}
\usepackage{amssymb}
\usepackage{epstopdf}

\newcommand{\Expect}[1]{\left\langle #1 \right\rangle}
\def\gsim{\mathrel {\vcenter {\baselineskip 0pt \kern 0pt \hbox{$>$} \kern 0pt \hbox{$\sim$} }}}
\def\p{\partial}
\def\ba{\begin{eqnarray}}
\def\ea{\end{eqnarray}}
\def\hf{\frac12}
\def\be{\begin{equation}}
\def\ee{\end{equation}}
\def\df{\delta\phi}

\def\lsim{\mathrel {\vcenter {\baselineskip 0pt \kern 0pt \hbox{$<$} \kern 0pt \hbox{$\sim$} }}}

\def\gsim{\mathrel {\vcenter {\baselineskip 0pt \kern 0pt \hbox{$>$} \kern 0pt \hbox{$\sim$} }}}
\def\p{\partial}

\title{Simple Bounds from the Perturbative Regime of Inflation}

\author{Louis Leblond$^1$ and Sarah Shandera$^2$\\
  $^1$ George P. \& Cynthia W. Mitchell Institute for Fundamental Physics\\
  Texas A\&M University, College Station, TX 77843-4242\\
\vskip .3cm 
$^2$Institute of Strings, Cosmology and Astroparticle Physics \\
Physics Department, Columbia University, New York, NY 10027\\
\vskip .3cm 
$^1$ \email{lleblond@physics.tamu.edu}, $^2$ \email{sarah@phys.columbia.edu}}
\abstract{We examine the conditions under which a perturbative expansion around an inflating background is valid. When inflation is driven by a  single field with a general sound speed, we find a lower limit on the sound speed related to the amplitude of the inflationary power spectrum.
Generalizing the sound speed constraints to include scale dependence can limit the number of e-folds obtained in the perturbative regime and restrict otherwise apparently viable models. We also show that for models with a low sound speed, eternal inflation cannot occur in the perturbative regime.}
\preprint{MIFP-08-1, arXiv:0802.2290}

\begin{document}

\section{Introduction}
Inflationary model building has been a popular pursuit in the years since data from the cosmic microwave background became precise enough to begin distinguishing scenarios. Model builders have taken either the approach of attempting to constrain the terms in a ``generic" model from observation (`reconstruction') or have searched for particular models with interesting features in the context of supersymmetry, string theory or related ideas. However, it is still not clear what range of observables, particularly for the amplitude of tensor fluctuations and non-Gaussianity, is reasonable or expected from fundamental theory. In addition, `reconstructors' often proceed without clearly stating or exploring the assumptions and context for the theories they are reconstructing. 

In either approach, there are very basic guidelines coming simply from the consistency of the perturbative approach to calculating fluctuations. 
The implicit assumption is that the inflaton obeys an effective field theory (often leaving aside the important question of vacuum selection, which we too will ignore here) and on scales below some cutoff its fluctuations can be calculated order by order. The conditions for this perturbative expansion to be valid are trivial and well known in the context of slow-roll. However, smooth, single-field slow-roll inflation cannot produce large non-Gaussianity \cite{Acquaviva:2002ud, Maldacena:2002vr}.
As an additional complication, single field slow-roll models are surprisingly hard to find in string theory and do not seem to contain features that shed light on their fundamental physics origin \cite{McAllister:2007bg, Hertzberg:2007wc}. Given all the interest in other models, then, it is interesting to explore the extension of the consistency constraints to other cases. Here we do so for models with small and changing sound speed like k-inflation \cite{Garriga:1999vw}, ghost inflation \cite{ArkaniHamed:2003uz} and Dirac-Born-Infeld (DBI) brane inflation \cite{Alishahiha:2004eh}. We also look at a toy example with multiple fields. This is a simple idea, and some of our results have been obtained in some form previously \cite{Cheung:2007st, Creminelli:2008es} so we emphasize the generality of the approach and some interesting applications. 

In the standard picture of inflation, the classical behavior of a scalar field drives the background accelerating expansion of the universe and quantum fluctuations of the field during inflation are eventually imprinted as perturbations in the gravitational potential after inflation. There are two basic consistency conditions here: first, that the energy density of fluctuations is not so large that it overwhelms the background, and second that interactions of the field are small enough that the amplitude of fluctuations can be well calculated from the term in the action quadratic in fluctuations. We will call the first condition the gradient energy condition, and in slow roll it is satisfied if $H/M_p<1$ which is anyway a good condition that keeps the energy density driving inflation sub-Planckian. 
\footnote{The gauge invariant method to calculate backreaction effects was discussed in \cite{Mukhanov:1996ak} and we thank the referee for pointing this out. 
Note that the condition we give in this paper 
does not include the stochastic behavior of the field and therefore is strongest near the onset of inflation. In that sense it is a minimal condition.} The second condition (the interaction picture constraint) is never violated in slow-roll as long as $H/M_p<1$ and the slow-roll parameter $\epsilon$ is less than one (also the condition for accelerated expansion). 

Particularly as we become interested in models that have large interactions generating large non-Gaussianity, it is useful to put these constraints in the slightly more formal, but also most general, language of the action. Schematically, the action for a general scenario can be written in powers of the fluctuations
\be
S=S_0+S_2+S_3+\dots
\ee
where $S_0$ describes the classical evolution, $S_1$ is absent when the background fields satisfy their equations of motion, $S_2$ is the quadratic piece that gives the two-point function (the power spectrum), and higher orders are interaction terms. Then the two conditions above can be phrased schematically 
\ba
S_0&>&S_2\\\nonumber
S_2&>&S_n,\; n\geq3
\ea
where the amplitude of terms can be estimated using the the average amplitude of fluctuations near horizon crossing. Since calculating higher orders in perturbation theory is a long chore, we will focus on the first case of the second line above, $S_2>S_3$. This is really a statement about a loop contribution to the two-point function, where we have suppressed a log factor.

The main application of these ideas in this paper is a lower bound on the sound speed for which the perturbative expansion is valid. This bound is particularly interesting since it imposes a strong constraint on models like DBI inflation, where the sound speed can vary considerably.  Given the sound speed at some scale $k_0$, we can compute the number of e-folds obtained in the perturbative regime, as a function of the running.  This also has implications for bounding the expected non-Gaussianity on non-CMB scales, where there is as yet no interesting observational constraint. This bound coincides with the expression found in \cite{Cheung:2007st} by an alternate method. This bound also implies that eternal inflation is outside of the pertubative regime in any model with small sound speed, which was realized independently in a recent paper of Creminelli et al \cite{Creminelli:2008es}. Finally we also find a bound on the number of fields, $N$, in a simplistic multiple field example which agrees completely with recent bounds from renormalization of the Planck mass \cite{Veneziano:2001ah} and from black hole physics \cite{Dvali:2007hz} applied in de Sitter space \cite{Dvali:2008sy}.

In the next section we review the basic framework of the perturbative calculation. In Section \ref{bounds} we examine in detail the regime of validity of the calculation in the case of slow-roll and general sound speed models with scale-dependence. Implications of the bounds for scale-dependent models are discussed in Section \ref{scaleDepend}. Multi-field inflation is discussed in Section \ref{multifield} and Section \ref{conclude} concludes.

\section{Set-up, Gauges and Conventions}
\label{setup}
Here we outline the class of theories we consider and choose a gauge for the calculation.
We will first consider the following system
\ba
\label{generalsinglefield}
S = \frac12\int d^4 x \sqrt{-g}\left(M_p^2 R + 2P(X,\phi)\right)
\ea
where $\phi$ is the inflaton and $ X = -\hf g^{\mu\nu}\p_\mu\phi\p_\nu\phi$. In usual slow-roll, $P(X,\phi)=X-V(\phi)$. We are interested in an inflationary background where the inflaton is homogeneous with perturbations $\phi(\vec{x}, t) = \phi_0(t) + \delta\phi(\vec{x},t)$. The background metric is the usual Friedmann-Robertson-Walker (FRW) metric
\ba
ds^2 = -dt^2 + a^2(t) \delta_{ij}dx^idx^j
\ea
Einstein's equation together with the continuity equation completely describe the (inflating) background
\begin{align}
 H^2 & =  \frac{\rho}{3M_p^2}\\
\dot\rho &= -3(\rho +P)
\end{align}
where $\rho$ is the energy density, $P$ is the pressure and $H = \frac{\dot a}{a}$ is the Hubble parameter. It is useful to define the speed of sound:
\ba
\label{cs}
c_s^2 = \frac{dP}{d\rho} = \frac{P_{,X}}{P_{,X} + 2XP_{,XX}}
\ea
and the ``slow-roll" parameters
\ba\label{slow-roll}
\epsilon &=& -\frac{\dot H}{H^2} = \frac{X P_{,X}}{M_p^2H^2}\nonumber\\
\eta &=& \frac{\dot\epsilon}{\epsilon H}\nonumber\\
s &=& \frac{\dot c_s}{c_s H}
\label{flowparam}
\ea
where $P_{,X}$ means the derivative of $P$ with respect to $X$. To analyze the validity of the perturbative analysis, we will work (following \cite{Maldacena:2002vr}) in the ADM formalism, where the metric is
\cite{Arnowitt:1962hi}:
\ba
ds^2 = -N^2dt^2 + h_{ij}(dx^i + N^idt)(dx^j + N^j dt)
\ea
The lapse functions, $N$, $N^{i}$ function as Lagrange multipliers which can be eliminated through their equations of motion.
Since the metric also fluctuates, it is important to choose a gauge and correctly account for all the fluctuations. In this paper, we will be using two different gauges, and we write only the scalar part:
\begin{enumerate}
\item In the comoving gauge, $\delta\phi = 0$ and $h_{ij} = a^2 e^{2\zeta} \delta_{ij}$.  One can solve for the lapse functions in this gauge \cite{Maldacena:2002vr}  to find to first order $N^i = \p_i \left(-a^{-2} \frac{\zeta}{H} + \p^{-2}\left(\frac{\dot\phi^2\dot\zeta}{2H^2c_s^2}\right)\right)$  and $N = 1 + \frac{\dot\zeta}{H}$.  $\zeta$ is constant outside of the horizon and corresponds to the measured curvature perturbation in the post-inflationary era.

\item In the spatially flat gauge we set $h_{ij} = a^2 \delta_{ij}$ and expand the action in terms of $\delta\phi$.  The lapse functions are then $N = 1 + \frac{\dot\phi}{2H} P_{,X} \delta\phi$ and $N^i = \p_i \p^{-2} \frac{a^2 \epsilon}{c_s^2} \frac{d}{dt}\left(-\frac{H}{\dot\phi}\delta\phi\right)$ \cite{Arroja:2008ga}.  In this gauge the amplitude of fluctuations of the inflaton is not constant after horizon exit and will change.
\end{enumerate}

One can move from one gauge to the other by a gauge transformation which to first order in the fluctuations is
\ba\label{gaugerelation}
\zeta = - \frac{H}{\dot\phi}\delta\phi
\ea
Working in the comoving gauge, we can expand the action in fluctuations of the metric $\zeta$. The first order vanishes since it is proportional to the background field's equation of motion while at second order
\ba
\label{S2}
S_2 = M_p^2\int dt d^3x \left(a^3\frac{\epsilon}{c_s^2}\dot\zeta^2 - a \epsilon (\p\zeta)^2\right)
\ea
Terms proportional to three powers of $\zeta$ ($S_3$) were calculated explicitly in \cite{Maldacena:2002vr} for usual slow-roll and for general sound speed in \cite{Seery:2005wm, Seery:2005gb, Chen:2006nt}. The full expression is quite long but contains the following significant terms:
\ba\label{theS3term}
S_3 = M_p^2\int dt d^3x \frac{a\epsilon}{c_s^2} (\epsilon-2s+1-c_s^2)\zeta (\p\zeta)^2 + \dots
\ea
The last three terms in the parentheses cancel for a canonical kinetic term where $c_s = 1$ and $s=0$, and the first term is the usual slow-roll piece which carries factors of $\epsilon^2$. 

\subsection{The Variance of $\zeta$}
While the system is stable at the classical level, quantum perturbations will give a non-zero variance to the random $\zeta$ perturbations. In calculating this variance, there are two assumptions involved:
\begin{enumerate}
\item The FRW background is valid and the correction from the perturbations is small compared to the background.
\item The dynamics of the $\zeta$ field is dominated by its quadratic action and higher order interactions are small.
\end{enumerate}
Therefore, as usual, one solves for the perturbation $\zeta$ using $S_2$.  The perturbations are decomposed into momentum modes
\ba
u(\vec{k}, t) = \int d^3x \zeta(\vec{x}, t) e^{-i\vec{k}\cdot\vec{x}}
\ea
and $\zeta$ is quantized
\be
\zeta(\vec{k}, t)=u(\vec{k},t)a(\vec{k})+u^*(-\vec{k},t)a^{\dag}(-\vec{k})
\ee
where as usual $[a(\vec{k}),a^{\dag}(\vec{k^{\prime}})]=(2\pi)^3\delta^3(\vec{k}-\vec{k^{\prime}})$.
The solution to the equation of motion, assuming all three slow-roll parameters above are small ($\epsilon$, $\eta$, $s\ll1$), is
\ba
u_k = \frac{-iH}{M_p\sqrt{4\epsilon c_s k^3}} (1+ ikc_s\tau) e^{-ikc_s\tau}
\ea
where $dt = a d\tau$ is the conformal time ($\tau \sim -\frac{1}{aH}$ for $\epsilon\ll1$). Before horizon exit $c_s k > aH$ and $u_k$ oscillates with a decreasing amplitude ($\propto \tau$). After horizon exit $|u_k|$ is constant and on the order of
\ba
u_k \sim \frac{H}{2M_p\sqrt{\epsilon c_s k^3}}
\ea
The two-point function is defined 
\ba
\Expect{\zeta(\vec{k}_1) \zeta(\vec{k}_2)} &\equiv& (2\pi)^3\delta^3(\vec{k}_1+\vec{k}_2) P_\zeta\\ \nonumber
&=&(2\pi)^3\delta^3(\vec{k}_1+\vec{k}_2)2\pi^2\mathcal{P}_{\zeta}k^{-3}
\ea
where $\mathcal{P}_{\zeta}$ is the dimensionless power spectrum (variance), and assuming the spectral index doesn't run $\mathcal{P}_{\zeta}=A(k_0)(k/k_0)^{n_s-1}$. Using the solution above, we also have
\ba
 \Expect{\zeta(\vec{k}_1) \zeta(\vec{k}_2)}&=&(2\pi)^3\delta^3(\vec{k}_1+\vec{k}_2)|u(\vec{k})|^2\\\nonumber
&=& (2\pi)^3\delta^3(\vec{k}_1+\vec{k}_2)\left(\frac{H^2}{4M_p^2\epsilon c_s}\right)k^{-3}
\ea 
Averaging over a sound horizon volume, $\zeta(\vec{x}, t)$ can be thought as a random variable with mean zero and fluctuations of size $\frac{H}{2\pi M_p\sqrt{2\epsilon c_s}}$. 
Note that a smaller sound speed means a smaller horizon for scalar modes. 
From the variance of $\zeta$ we can do a gauge transformation and compute the variance of $\delta\phi$, we find 
\ba
\label{phifromzeta}
\Expect{\df^2}^{1/2} = \frac{|\dot\phi|}{H} \Expect{\zeta^2}^{1/2} \sim \frac{H}{2\pi \sqrt{c_s P_{,X}}}
\ea
where we have used the definition of $\epsilon$ in Eq.(\ref{slow-roll}) and $2X = \dot\phi^2$. For a canonical kinetic term the variance is of order $H^2/4\pi^2$.  Now let us check under what conditions our two basic assumptions are valid.  For the rest of this paper, we will be mostly interested in order of magnitude calculations and we will drop the factors of $2\pi$ and take $\zeta \sim \frac{H}{M_p \sqrt{\epsilon c_s}}$ over scales of length $L \sim \frac{c_s}{aH}$ and time of order $1/H$.

\section{Perturbative Limits During Inflation}
\label{bounds}
The first condition one must have for a valid perturbative analysis of inflation is that the background is stable. That is, we demand that the contribution to the energy density coming from the pertubations is smaller than the smooth constant energy density that drives inflation. This is ensured by demanding that $S_2 < S_0$.  For an inflating solution where the potential dominates, one gets that $S_0 \sim V(\phi_{(0)})$ which is  of order $H^2 M_p^2$ using the Friedmann equation.  Given that $\zeta\sim\frac{H}{M_p\sqrt{\epsilon c_s}}$ over a Hubble scale with
\begin{align}
\label{roughderivs}
\frac{\p}{\p t} \sim H\; , & & \frac{\p}{\p x} \sim \frac{a H}{c_s}\; ,
\end{align}
the gradient energy bound for a general single field model is
\begin{align}
\frac{S_2}{S_0} &\sim \frac{\epsilon a^{-2}(\p\zeta)^2}{H^2 M_p^2}\\
&\sim \frac{H^2}{M_p^2 c_s^3} < 1
\end{align}
The kinetic energy term in $S_2$ contributes at the same order. Note that for usual slow-roll models with $c_s =1$, this condition just reduces to having a subplanckian Hubble scale.  When $c_s$ is small the gradient energy induced from the perturbations is enhanced in two ways: the variance increases while the horizon decreases. We therefore find that $H$ must be significantly less than the Planck scale in a small sound speed model.  
\be
\frac{H^2}{M_p^2} < c_s^3
\label{bound1}
\ee
This condition can also be stated in terms of the stress-energy tensor, using the spatially flat gauge. The gradient energy condition is
\ba
\frac{\Expect{\delta T_0^0}}{T_0^0}& < &1
\ea
The first non-zero contribution to $\Expect{\delta T_0^0}$ contains the same terms (kinetic energy and gradient energy of fluctuations) as the action at second order in fluctuations, $S_2$ given in Eq.(\ref{S2}). For potential energy dominated inflation, we can shorthand $T_0^0$ as $3M_p^2H^2\approx V(\phi)$, which is also the dominant term in $S_0$.

\subsection{The Expansion Parameter to Third Order}
While the first condition is somewhat trivial, the condition demanding that all the higher order interaction terms are small is more interesting. Higher order terms in the action will correct the two point function. An example loop calculation was done in \cite{Weinberg:2005vy} (for a case with multiple fields, which we return to in Section \ref{multifield}) with the conclusion that, up to a term logarithmic in scale $k$, $S_3/S_2\ll1$ is required for the loop contribution to be small. The importance of the log term is considered in detail in \cite{Seery:2007we, Seery:2007wf, Sloth:2006az, Sloth:2006nu}, but we emphasize that the simplest conditions must hold when the log is of order 1. Although for a generic model the loop calculation is quite involved, the bound that results is not surprising. The well-studied local model, for example, parametrizes non-Gaussianity in a simple way that helps us see what to expect. In the local model the primordial curvature perturbation in real space is written as a Gaussian piece, $\zeta_g$, plus a quadratic correction
\be
\zeta=\zeta_g+\frac{3}{5}f_{NL}(\zeta_g^2-\Expect{\zeta_g^2})
\ee
A quick computation shows that the two-point function is not significantly corrected while $f^2_{NL}\Expect{\zeta_g^2} \ll1$ and in that case the fluctuations are nearly Gaussian. For fluctuations generated during inflation, `nearly Gaussian` means that the primary inflaton field is nearly free. In single field slow-roll this statement is related to the flatness of the potential. More generally, `nearly Gaussian' means that we can solve the quadratic equation of motion for the fluctuations and treat interactions as perturbations, that is\footnote{A quick, approximate evaluation of the correlation functions can be made assuming the dominant contribution comes from one Hubble time near horizon exit. Then $\Expect{\zeta^n}\approx\Delta t(\Delta x)^3\mathcal{L}_2(\mathcal{L}_n/\mathcal{L}_2)\mathcal{P}_\zeta^{n/2}$, where $\mathcal{L}_n$ is the Lagrangian at $n$th order in fluctuations. Using Eq.(\ref{S2}) and Eq.(\ref{roughderivs}), the product of the first three terms is about 1, so the dimensionless quantity $\Expect{\zeta^n}/\mathcal{P}_{\zeta}^{n/2}\approx \mathcal{L}_n/\mathcal{L}_2$, which is the same as $S_n/S_2$ as we are using it. The same dimensionless combination of the $n$-point function scaled by the two-point appears in the expansion when one writes the probability distribution function as a Gaussian modified by a series of cumulants.} $S_n/S_2\ll1$.  

With these examples in mind, we assume that the requirement $S_3/S_2\ll1$ also follows from the loop calculations with the terms we are interested in. Requiring $S_3<S_2$ and using the 1-sigma size of $\zeta$ gives
\begin{align}
\frac{S_3}{S_2}&\sim \frac{\zeta}{c_s^2} (\epsilon - 2s + 1 - c_s^2)  <1\nonumber \\
 &\sim  \frac{H}{c_s^2M_p (\epsilon c_s)^{1/2}}(\epsilon  - 2s + 1 - c_s^2) <1
\end{align}

Depending on whether the model in question is a slow-roll case (where $c_s = 1$ and $s = 0$) or is a small sound speed case ($1-c_s^2 >\epsilon - 2s$) there are two different bounds:
\begin{align}
&\frac{H^2\epsilon}{M_p^2} < 1 & \text{slow-roll}\\
&\frac{H^2}{M_p^2\epsilon} <c_s^{5} & \text{small}\; c_s
\end{align}
When the sound speed is small and $c_s<1/\epsilon$ (always true during inflation), this bound is more constraining than the gradient energy bound. This last equation is particularly interesting since one can write it in terms of the observed magnitude of the curvature power spectrum to find a lower bound on the sound speed
\begin{align} \label{bound2}
c_s^4 & > \frac{H^2}{M_p^2\epsilon c_s}\sim \mathcal{P}_\zeta.
\end{align}
This gives a direct experimental bound on how small $c_s$ was when CMB scales left the horizon. Indeed, Cosmic Microwave Background observations indicate $\mathcal{P}_\zeta \sim10^{-9}$ \cite{Spergel:2006hy}, implying $c_s > 10^{-9/4}$ for a perturbative analysis of inflation to make sense at CMB scales. This relation agrees with the bound found in \cite{Cheung:2007st} by a different method.

The bound on $c_s$ translates into a (somewhat model dependent) bound on the magnitude of non-Gaussianity coming from the three-point function: $|f^{eff}_{NL}| \sim \frac{1}{c_s^2}  < 10^{9/2}$. Here $|f^{eff}_{NL}|$ is calculated at the equilateral limit ($k_1=k_2=k_3$) of the momentum-space three-point function. The bound again agrees with our expectations from the local model. This is much weaker than current experimental constraints at CMB scales, which give roughly $|f^{eff}_{NL}| < 300$ for small sound speed models \cite{Creminelli:2006rz}. However, these bounds must be satisfied over the full duration of inflation in order to be able to trust the perturbative analysis. For models where the sound speed decreases during inflation, we will use this result to derive a constraint on $c_s$ in terms of the required duration of the inflationary phase. In particular DBI brane inflation has a sound speed that decreases during inflation as the brane moves along a space with continually decreasing warp factor.

We have concentrated on some particular terms in $S_3$ (largest when $c_s$ is small), but there are certainly others. In fact, there are two families of terms that introduce new parameters at each order in the expansion of the action. The first family comes from interaction terms in the potential and can be arranged in terms of the Hubble slow-roll hierarchy parameters (always assuming the potential energy dominates) \cite{Liddle:1994dx}, where new parameters enter that are related to consecutive derivatives of $\epsilon$. The second is an analogous family coming from the kinetic terms. For example, at third order it is useful to define
\ba
\lambda=X^2P_{,XX}+\frac{2}{3}X^3P_{,XXX}=\frac{\epsilon H^2}{c_s^2}\left[\frac{1}{6}\left(\frac{1}{c_s^2}-2\right)-\frac{2Xc_{s,X}}{3c_s}\right]
\ea
From the reconstruction point of view, there is a constraint on the new parameters at each order. For example, the term in $S_3$ containing $\lambda$ is of the order $\frac{2\lambda H^3}{\epsilon^{3/2} c_s^{3/2}}$ and this must also be less than the contribution from $S_2$:
\ba
\frac{\lambda c_s^{3/2}}{\epsilon^{3/2}} < HM_p
\ea
Note that throughout this section, we have assumed there is no magical cancellation among terms at any order. From an effective field theory viewpoint, such cancellations would require fine-tuning.

\subsection{Implications for Eternal inflation}

Eternal inflation will occur in any region of space where the magnitude of the 
density perturbation
\begin{equation}
\mathcal{P}_\zeta \sim \frac{H^2}{M_p^2\epsilon c_s}
\end{equation}
is of order 1, which requires curvature perturbations of order 1 (see \cite{Creminelli:2008es} for a more precise definition). For single field slow-roll inflation, this statement comes from demanding that the quantum perturbations of the inflaton $\delta\phi_q \sim H$ are of the same order as the classical motion $\delta\phi_c \sim \frac{|\dot\phi|}{H}$. This is clear since using Eq. (\ref{gaugerelation})
\be
\zeta = -\frac{H}{\dot\phi} \delta\phi_q  \sim \frac{ \delta\phi_q}{\delta\phi_c}
\ee
Now interestingly the bound (\ref{bound2}) implies 
\be
\mathcal{P}_\zeta < c_s^4 < 1
\ee
and therefore eternal inflation is not perturbatively possible in this class of models.  It has been claimed in the literature that eternal inflation should actually be more common in the phase space of configuration for models with small sound speed. This was observed in \cite{Helmer:2006tz, Tolley:2008na} based on the stochastic approach to inflation and was attributed to the increase of the diffusion coefficient thanks to the small sound speed. We have shown here that actually all that eternally inflating part of the phase space is outside of the perturbative regime and the two-point function cannot be calculated reliably using perturbative methods. In the specific case of DBI inflation in string theory, it has been verified that for the parameter range allowed by the theory, eternal inflation does not occur \cite{Chen:2006hs}. We emphasize that our statement does not rule out eternal inflation in small sound speed model. Instead, we are saying that such an eternally inflating regime will always be outside of the perturbative regime.
Note that the bound (\ref{bound2}) is only valid for small $c_s$. In the limit where $c_s = 1$ the large terms leading to (\ref{bound2}) vanish as $(1-c_s^2$) and one is left with the terms that are supressed by an extra power of $\epsilon$ giving
\be
\mathcal{P}_\zeta < \frac{1}{\epsilon^2}
\ee
and clearly for small enough $\epsilon$, eternal inflation is possible in the slow-roll $c_s = 1$ perturbative regime. While slow-roll eternal inflation can be perturbative locally, there might still be regions of the eternally inflating universe where the curvature of space is greater than $1/\epsilon$. Indeed, a global picture of an eternally inflating universe most likely requires a non-perturbative description even in the slow-roll picture \cite{Goncharov:1987ir}.

\subsection{Higher Order Terms}
\label{higher}

In the previous subsection, we found the conditions for $S_3<S_2$ and $S_2<S_0$.  These conditions matched our physical expectation in the simplest slow-roll cases, and we expect that we have uncovered the `expansion' parameters.  The calculation to fourth order has been recently pursued by different groups. To get the action to fourth order in the perturbation, one needs to expand the lapse functions $N$ and $N^i$ to second order in the perturbations (see \cite{Chen:2006nt} for more details on this point). This calculation for the action (\ref{generalsinglefield}) in the slow-roll limit was carried by Seery, Lidsey and Sloth \cite{Seery:2006vu} who found that the leading terms (like the gradient terms in $S_2$) are not suppressed by the slow-roll parameters and $S_4$ scales like $H^6/M_p^2$. More recently, Arroja and Koyama \cite{Arroja:2008ga} have argued that tensor modes must also be included and give a contribution of the same order \footnote{We thank F. Arroja and K. Koyama for clarifying this point to us.}. We see that $S_4$ will be small once the two conditions uncovered in the last section are satisfied. While the terms na\"ively order themselves correctly in powers of $\zeta$, a strong scale dependence of the parameters can change that unless a condition on the coefficients appearing in the action is satisfied. As we will show, this can be particularly troublesome in models with scale dependent sound speed.

In the small sound speed limit, the leading terms are coming from the self-interaction of the inflaton and one can neglect the coupling to gravity \cite{Huang:2006eh, Arroja:2008ga}. Since the slow-roll scaling of terms with powers of $\epsilon$ is clearest in the spatially flat gauge, where the fluctuations of the inflaton field, $\phi=\phi_{(0)}+\delta\phi$, are non-zero we will work in this gauge in this section.  
We can then expand $P(X,\phi)$ in terms of $\df$. We will not uncover all the terms - in particular, we will not display terms proportional to metric fluctuations in this gauge, which enter $S_3$ and higher. The purpose here is to show the behavior of the terms which, from experience so far, give the dominant contribution at each order. We will use the result as a starting point to discuss the case of scale-dependent sound speed. We switch to Lagrangian notation to avoid carrying around integral signs. 

We examine the kinetic terms, finding at each order new combinations of higher derivatives. For reference, two previously defined useful combinations are
\ba
\label{SigLam}
\Sigma&=&XP_{,X}+2X^2P_{,XX}=\frac{XP_{,X}}{c_s^2}=\frac{\epsilon H^2M_p^2}{c_s^2}\\\nonumber
\lambda&=&X^2P_{,XX}+\frac{2}{3}X^3P_{,XXX}
\ea
Notice that $X$ and $P_{,X}$ are positive and $c_s>1$ would mean $P,_{XX}<0$. As discussed in \cite{Adams:2006sv}, there is good reason to believe that this is not consistent with a unitary effective field theory. A useful expression is
\be
\label{csPxx}
\frac{1}{c_s^2}-1=\frac{2XP_{,XX}}{P_{,X}}
\ee
Following \cite{Huang:2006eh},  we can Taylor expand the action for the inflaton $P(X,\phi)$ in terms of the perturbation $\df$. For brevity, we keep only the $P_{X}$ derivatives, which give a larger contribution in models with small and slowly changing sound speed anyway:
\ba
a^{-3}\mathcal{L}_2&=&\frac{1}{2}P_{,X}[\dot{\df}^2-a^{-2}(\nabla\df)^2]+\frac{1}{2}\dot{\phi}^2\dot{\df}^2P_{,XX} + \dots\\\nonumber
a^{-3}\mathcal{L}_3&=&\frac{1}{2}P_{,XX}\dot{\phi}\dot{\df}[\dot{\df}^2-a^{-2}(\nabla\df)^2] +\frac{1}{6}P_{,XXX}\dot{\phi}^3\dot{\df}^3 +\dots\\\nonumber
a^{-3}\mathcal{L}_4&=&\frac{1}{8}P_{,XX}[\dot{\df}^2-a^{-2}(\nabla\df)^2]^2+\frac{1}{4}P_{,XXX}\dot{\phi}^2\dot{\df}^2[\dot{\df}^2-a^{-2}(\nabla\df)^2]\\\nonumber
&&+\frac{1}{24}P_{,XXXX}\dot{\phi}^4\dot{\df}^4+\dots
\ea
Inserting powers of $\dot{\phi}^2=2X$ (the background spatial gradient is negligible), and using Eq.(\ref{phifromzeta}) for the variance of $\df$ and to label the combination $\mathcal{P}_{\zeta}$ gives
\ba\label{higher}
\mathcal{L}_2&=&a^3\frac{H^4}{c_sP_{,X}}\left[P_{,X}\left(1-\frac{3}{c_s^2}\right)+2XP_{,XX}\right]+ \dots=-2a^3\Sigma \mathcal{P}_{\zeta}+\dots\\\nonumber
\mathcal{L}_3&=&a^3\frac{1}{2}\left(\frac{2H^4}{\dot{\phi}^2c_sP_{,X}}\right)^{3/2}\left[X^2P_{,XX}\left(1-\frac{3}{c_s^2}\right)+\frac{2}{3}X^3P_{,XXX}\right] +\dots\\\nonumber
&&=-\mathcal{L}_2\frac{P^{1/2}_{\zeta}}{2c_s^2}\left[\frac{2\lambda c_s^2}{\Sigma}-3(1-c_s^2)\right]\\\nonumber
\mathcal{L}_4&=&\frac{a^3}{2}P^2_{\zeta}\left[X^2P_{,XX}\left(1-\frac{3}{c_s^2}\right)^2+4X^3P_{,XXX}\left(1-\frac{3}{c_s^2}\right)\right.\\\nonumber
&&\left.+\frac{4}{3}X^4P_{,XXXX}\right]+\dots
\ea
From the last line, we can see that although there will be a new parameter involving $P_{,XXXX}$, the first contribution will scale like $\mathcal{L}_2\mathcal{P}_{\zeta}/c_s^4$ where $\mathcal{P}_\zeta \sim \frac{H^2}{M_p^2\epsilon c_s} \sim \frac{H^4}{XP_{,X}c_s}$.
For the special case of DBI, the pressure is given by
\be
P^{DBI}(X,\phi)=-f(\phi)^{-1}\sqrt{1-2Xf(\phi)}+f(\phi)^{-1}-V(\phi)
\ee
where $f(\phi)$ depends on the higher dimensional background the brane is moving in. Then
\ba
c_s^2&=&1-2Xf(\phi)\\\nonumber
P_{,X}&=&\frac{1}{c_s}\\\nonumber
P_{,XX}&=&\frac{f(\phi)}{c_s^3}\\\nonumber
P_{,X^n}&= &(2n-3)!! \left[\frac{(P_{,XX})^{n-1}}{(P_{,X})^{n-2}}\right]\; ,n\geq3
\ea
with $(2n-3)!! = (2n-3)(2n-5)\cdots 1$. Using Eq.(\ref{SigLam}) and Eq.(\ref{csPxx}) in the last line gives
\be
X^nP_{,X^n}=(2n-3)!! \left(\frac{1}{c_s^2}-1\right)^{n-1}\frac{\Sigma c_s^2}{2^{(n-1)}}\; ,n\geq2
\ee
Then rewriting Eq.(\ref{higher}), we get (using the simplification for the DBI case that $2\lambda c_s^2/\Sigma=(1-c_s^2)$)
\ba
\mathcal{L}_2&\propto& \frac{H^4a^3}{2c_s^3}\\\nonumber
\mathcal{L}_3&\propto &\mathcal{L}_2L_3\frac{\mathcal{P}_{\zeta}^{1/2}}{c_s^2}\\\nonumber
\mathcal{L}_n&\propto &\mathcal{L}_2L_n\left(\frac{\mathcal{P}_{\zeta}^{1/2}}{c_s^2}\right)^{n-2}\\\nonumber
\label{ExpandL}
\ea
where the $L_n$ are constants. We remind the reader that this only considering the leading terms in the action for the small sounds speed and this is specific to DBI in the sense that we have assumed a specific form for the parameters $\lambda$ (and equivalent higher order generalizations).  Nevertheless, we expect the scaling argument of the next section to apply in principle to the general case.

\section{Scale Dependence}
\label{scaleDepend}
We can apply the bounds from the previous sections in an interesting way for models with scale-dependent sound speeds (which are observable as scale-dependent non-Gaussianity). Just as the Hubble parameter evolves slowly in time in a generic inflation model, many scenarios may also have a changing sound speed. In analogy to the spectral index (and using the parameter $s$ defined in Eq.(\ref{flowparam})), we can model this by writing
\be
\label{csscale}
c_s(k)=c_s(k_0)\left(\frac{k}{k_0}\right)^{s}
\ee
Let us suppose $s$ is a (nearly constant and small) parameter. If $s<0$, the sound speed will decrease on small scales (and so the magnitude of non-Gaussianity will increase). An early discussion of varying sound speed can be found in \cite{Chen:2005fe}. To connect more directly with observation, we can write the observable magnitude of the non-Gaussianity as
\be
\label{fnlscale}
f^{eff}_{NL}=f^{eff}_{NL}(k_0)\left(\frac{k}{k_0}\right)^{n_{NG}-1}
\ee
where the `effective' (eff) label indicates that this applies to more than the local model. Note also that this parameter $f_{NL}^{eff}$ is really an approximate amplitude for the full bispectrum that is evaluated in the equilateral limit where all momemta are equals.  In small sound speed models, $f^{eff}_{NL}\propto1/c_s^2$ so $n_{NG}-1=-2s$ and there is a model dependent coefficient relating $f^{eff}_{NL}(k_0)$ to $c^{-2}_s(k_0)$.

First, we derive a simple relationship between the number of e-folds obtained while the perturbative expansion is sensible at this level and apply it to the case of DBI inflation. Rewriting Eq.(\ref{bound2}) to display the scale-dependence gives
\ba
\mathcal{P}_\zeta&<&c_s^4\\\nonumber
A(k_0)\left(\frac{k}{k_0}\right)^{n_s-1}&<&c_s^4(k_0)\left(\frac{k}{k_0}\right)^{4s}\\\nonumber
\log\left[\frac{A(k_0)}{c_s^4(k_0)}\right]&<&(4s-n_s+1)\log\left[\frac{k}{k_0}\right]\\\nonumber
\ea
If $4s>0.05$, the sound speed increases with scale and the bound is satisfied for all $k>k_0$ (that is, for the duration of inflation). Note that $n_s - 1$ depends on $s$ in general \cite{ArmendarizPicon:2003ht}. Otherwise, for $H$ nearly constant, this says that the number of e-folds obtained before the bound is violated (or before $s$ or $n_s$ must change) is
\be
N_e^{max}=\frac{1}{(4s-n_s+1)}\log\left[\frac{A(k_0)}{c_s^4(k_0)}\right]
\ee
Contour plots below show this relation, assuming $n_s=0.95$ and $A_0=2.22\times10^{-9}$ at $k_0=0.002Mpc^{-1}$ \cite{Spergel:2006hy}, both in terms of initial sound speed $c_s(k_0)$ (Eq.(\ref{csscale})) and in terms of an effective magnitude $f^{eff}_{NL}$ (Eq.(\ref{fnlscale})), found using the dominant contribution in DBI evaluated at the equilateral triangle limit.
\begin{figure}[h]
\begin{center}
$\begin{array}{cc}
\includegraphics[width=0.5\textwidth,angle=0]{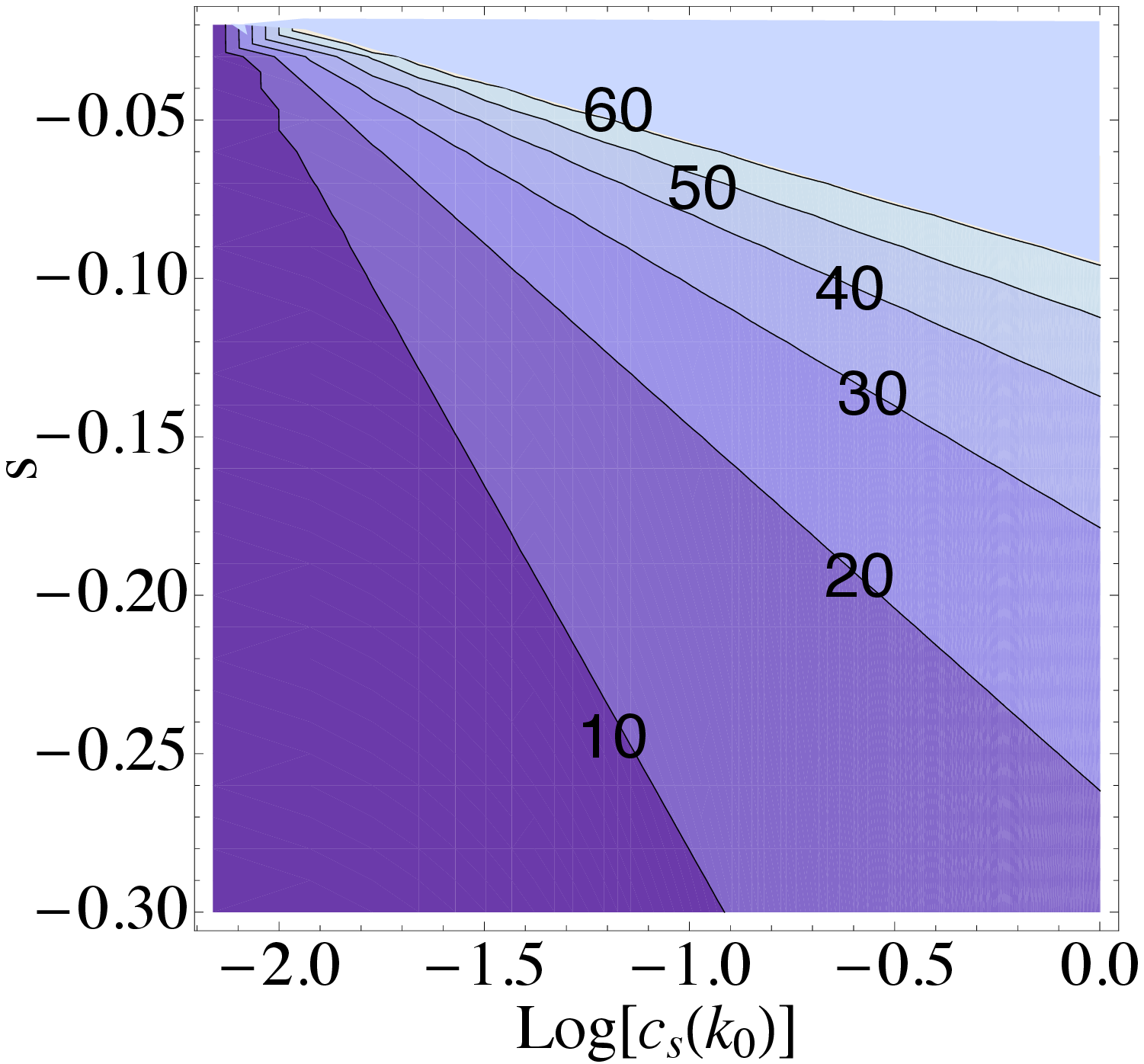} &
\includegraphics[width=0.463\textwidth,angle=0]{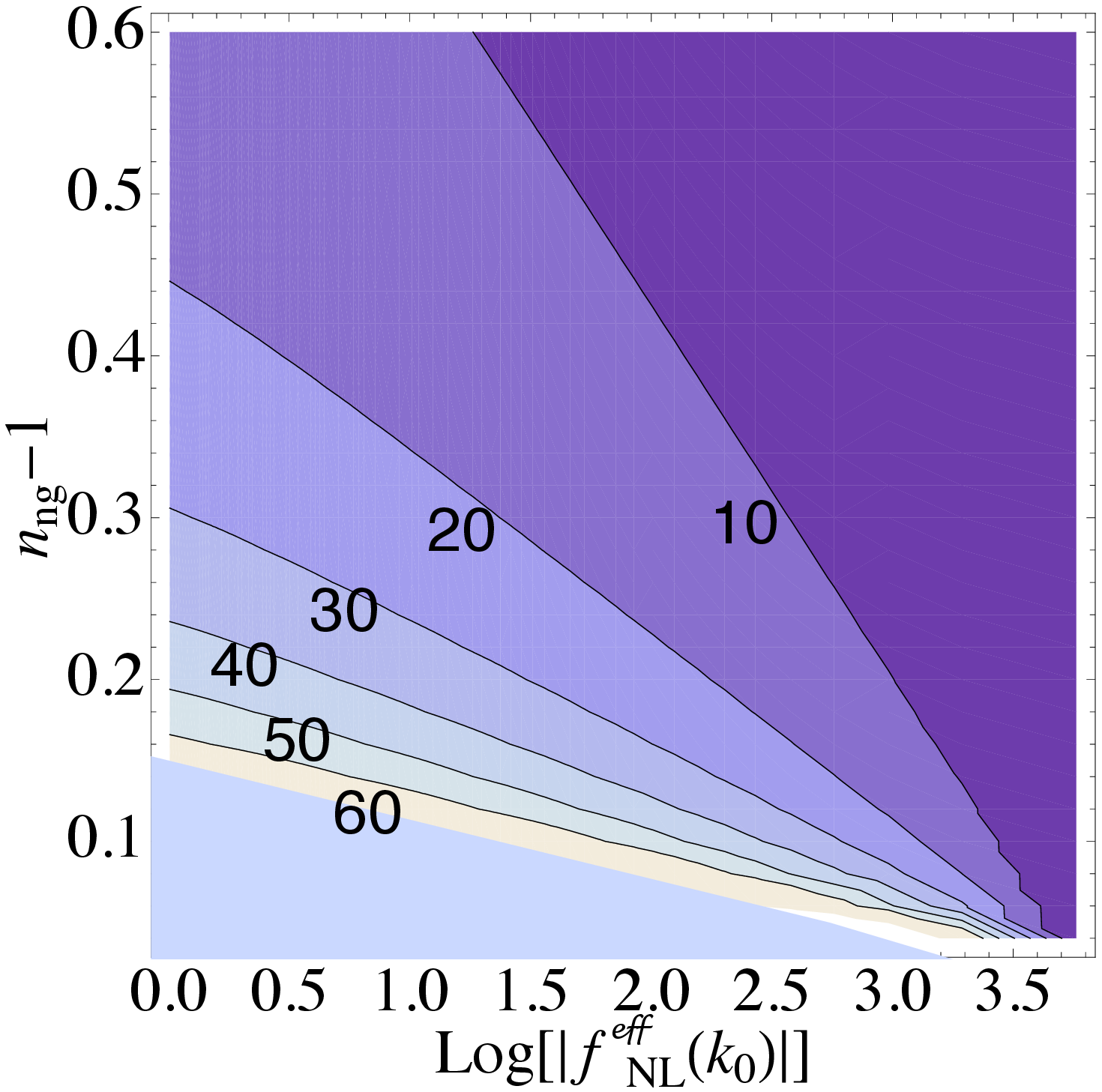} \\
\mbox{(a)} &\mbox{(b)}
\end{array}$
\caption{(a) Contours of number of e-folds obtained before the bound is violated, as a function of the sound speed at some initial scale $k_0$ and the scale-dependence $s$. (b) Contours of number of e-folds obtained in the DBI model before the bound is violated, as a function of $f^{eff}_{NL}\approx-0.32/c_s^2$ at scale $k_0$ and with scale-dependence $n_{NG}$.}
\label{Efoldscontour}
\end{center}
\end{figure}

The best case scenario for scale-dependent non-Gaussianity to be observable using upcoming data on cluster number counts and the galaxy bispectrum has both large non-Gaussianity on CMB scales and moderate running ($-s\sim0.1-0.3$) \cite{LoVerde:2007ri}. The plots above show that this case is consistent (since the scales differ by about a factor of ten, or a few e-folds) as long as $s\rightarrow0$ on scales just below the cluster scale, which is probably out of the easily observable range. 

In brane inflation where a brane falls down a warped throat, $s$ is roughly constant while the brane is in the AdS part of the throat. We can see from panel (b) of Figure \ref{Efoldscontour} that under these conditions with observable non-Gaussianity at CMB scales and running greater than about $n_{NG}-1 \sim0.2$, this model cannot achieve enough inflation and remain in the perturbative regime. However, there is a possible solution which suggests a solution for general sound speed models: the throat is not singular and in fact $s$ is zero in the smoothed out tip region where the warp factor is a constant. This feature has been analyzed previously in brane inflation models (for example, \cite{Kecskemeti:2006cg}). In order to get enough inflation in this scenario, one can try to find a particular choice of model parameters where $c_s$ does not get too small before entering the tip region. We can estimate the number of e-folds in the tip region in a simple case, taking $H=A\phi$ and using $\dot{\phi}=-2M_p^2H^{\prime}c_s$ (which can be derived from the continuity equation). Then
\ba
N_e^{tip}&=&\int_{\phi_t}^{\phi_f}\frac{H}{\dot{\phi}} d\phi=-\frac{1}{2M_p^2}\int_{\phi_t}^{\phi_f}\frac{H}{c_sH^{\prime}} d\phi\\\nonumber
&\approx&\frac{\phi_t^2}{4c_s(\phi_t)M_p^2}
\ea
where $c_s(\phi_t)$ is the (constant) value of $c_s$ in the tip region, $\phi_t$ is the value of $\phi$ at the edge of the tip region and we have assumed $\phi_f$ (the final value of the inflaton) is much smaller than $\phi_t$.
Since $\phi_t < M_p$ is required by the set-up, we see that the number of e-folds at the tip is bounded to be less than $N_e^{tip} < 1/4c_s(\phi_t)$ and in general it will be much smaller.
Then, the number of e-folds from the AdS part of the throat (e.g., read off the plot above for a particular case) and the tip region must together provide sufficient inflation. This can be imposed more precisely for specific scenarios and may be quite restrictive. 

We now turn to a more speculative condition on the coefficients of terms at each order. This is speculative in the sense that it assumes the terms calculated in the previous section have captured the scaling behavior of all of the most important terms that appear when the sound speed is small. In addition, it assumes that one is interested in (or expects) a non-Gaussian distribution which can be expanded around the Gaussian in a series of converging cumulants. While this is not a necessary feature of non-Gaussian distributions, it is commonly assumed by cosmologists and is likely to be true when the physics of the fundamental inflaton is simple (e.g., not eternally inflating).

If the background solution gives rise to a scale-dependent sound speed, that is if $s\neq0$, the terms at each order in the expansion also have a scale-dependence\footnote{See \cite{Lyth:2006gd,  Sloth:2006az, Sloth:2006nu} for a discussion of scale-dependence in slow-roll loop calculations.}. Explicitly writing the scale dependence of the leading term at each order in the Lagrangian (Eq.(\ref{ExpandL})) not dependent on $\lambda/\Sigma$ and other new parameters, we find 
\be
\hat{\mathcal{L}_n}\equiv\frac{\mathcal{L}_n}{\mathcal{L}_2}\propto L_n\left(\frac{\mathcal{P}_{\zeta}^{1/2}(k_0)}{c^2_s(k_0)}\right)^{(n-2)}\left(\frac{k}{k_0}\right)^{(\frac{n_s-1}{2}-2s)(n-2)}\equiv \hat{L}_n\left(\frac{k}{k_0}\right)^{q(n-2)}
\ee
for $n\geq3$ and where $q=(n_s-1)/2-2s$. Different factors of $2\pi$, etc, in the definition of $\mathcal{P}_{\zeta}$ will make no difference. Clearly, if $s<(n_s-1)/4\sim-0.01$, higher order terms will have a stronger scale dependence. How can we be sure, then, that the fluctuation Lagrangian expanded to order $n$ has actually captured the dominant contributions on scales smaller than $k_0$? This is relevant, for example, for calculations that involve the probability density function, which depends in principle on all correlation functions and is often treated as an expansion in moments around a Gaussian. We can demand that at any scale $k$, only a finite number of terms in the expansion contribute. Label $k_n^{*}$ the scale where $\hat{\mathcal{L}_n}=1$. Then if $k_n^{*}>k_{n-1}^{*}$ for all $n$, we can be sure that at scale $k_n^{*}$ the observables calculated from the Lagrangian expanded to order $n$ are correct. Then, we can write
\be
\hat{L}_n\left(\frac{k_n^{*}}{k_0}\right)^{q(n-2)}=\hat{L}_{n-1}\left(\frac{k_{n-1}^{*}}{k_0}\right)^{q(n-3)}
\ee
For the case $n=3$, this gives
\ba
\hat{L}_3\left(\frac{k_3^{*}}{k_0}\right)^{q}&=&\hat{L}_{2}=1\\\nonumber
\left(\frac{k_0}{k_3^{*}}\right)^{q}&=&\hat{L}_3
\ea
Then demanding $k_0/k_3^{*}<1$ also means that for $q>0$, $(k_0/k_3^{*})^q<1$, and so $\hat{L}_3<1$. But $\hat{L}_3=\mathcal{P}_\zeta^{1/2}(k_0)c^{-2}_s(k_0)$, so we have just recovered Eq.(\ref{bound2}). 

\begin{figure}[h]
\begin{center}
$\begin{array}{cc}
\includegraphics[width=0.5\textwidth,angle=0]{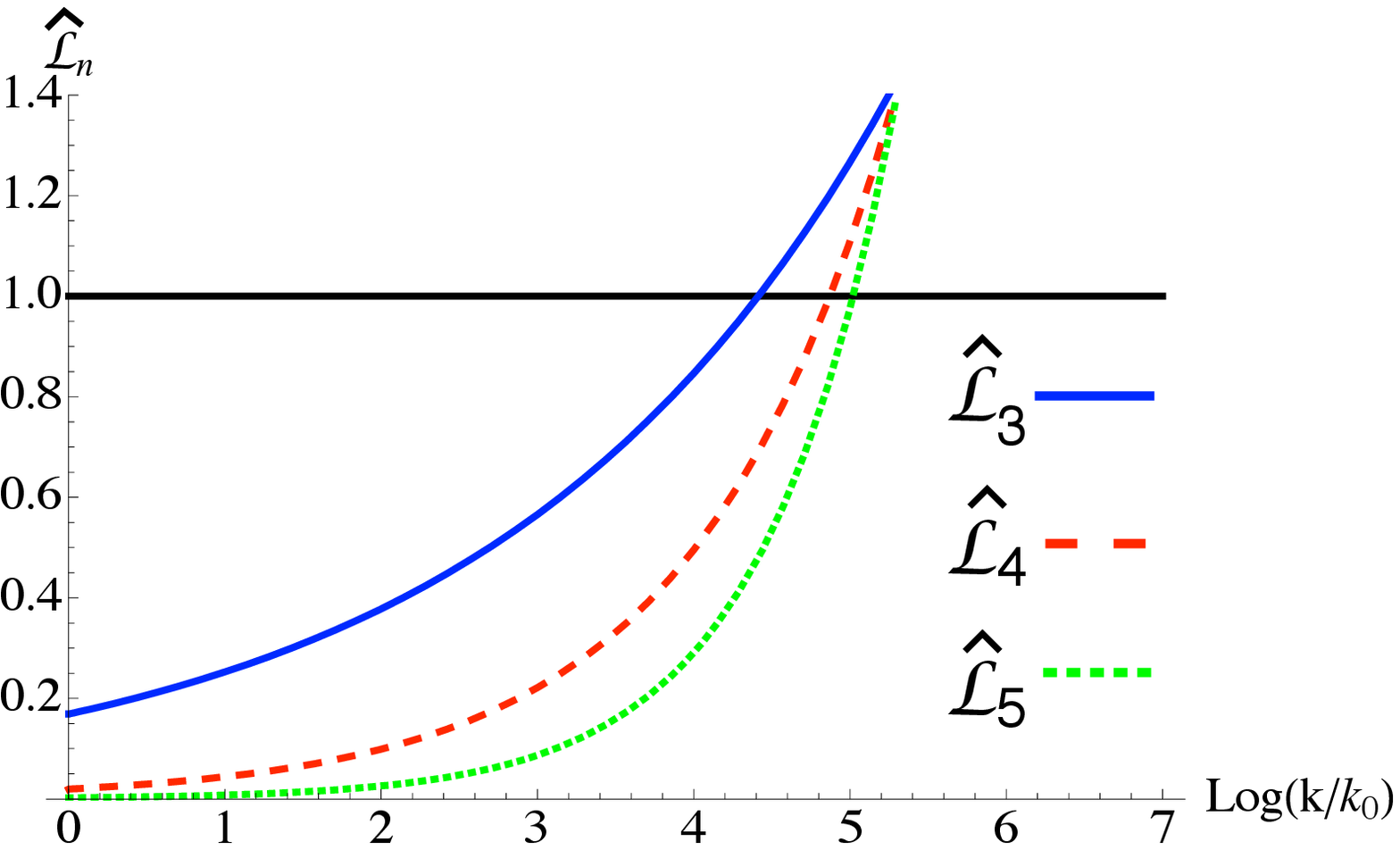} &
\includegraphics[width=0.5\textwidth,angle=0]{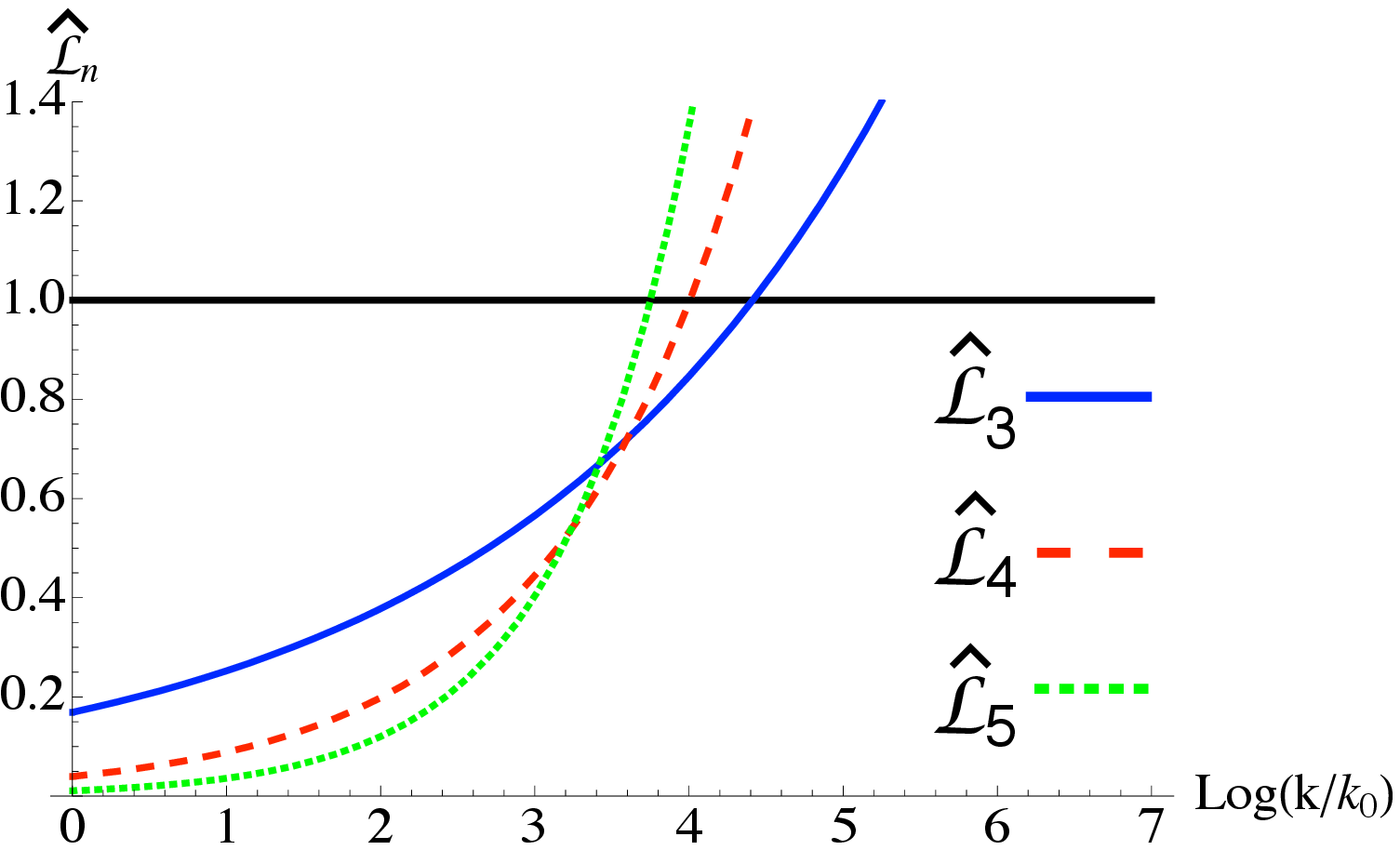} \\
\mbox{(a)} &\mbox{(b)}
\end{array}$
\caption{(a) Scaling of $\hat{\mathcal{L}}_3$ (solid blue), $\hat{\mathcal{L}}_4$ (long-dashed red), $\hat{\mathcal{L}}_5$ (short-dashed green) with coefficients that obey the bound in Eq.(\protect\ref{Lnbound}), with $s=-0.1$. (b) Same as before, but with coefficients that do not obey the bound. Note that when the equality is saturated, all terms would meet at the point where $\hat{\mathcal{L}}_3$ crosses 1.}
\label{scaling}
\end{center}
\end{figure}

For $n>3$, we have
\ba
\left(\frac{k_n^{*}}{k_0}\right)&=&(\hat{L}_n)^{-[q(n-2)]^{-1}}\\\nonumber
\left(\frac{k_{n-1}^{*}}{k_n^{*}}\right)^{q(n-2)}&=&\left(\frac{1}{\hat{L}_{n-1}}\right)^{(n-2)/(n-3)}\hat{L}_{n}
\label{relateLn}
\ea
Again, the left hand side must be less than one, so finally we have
\be
\hat{L}_{n}\leq(\hat{L}_{n-1})^{(n-2)/(n-3)}\Rightarrow L_n\leq(L_{n-1})^{(n-2)/(n-3)}
\label{Lnbound}
\ee
The implications of this result are shown graphically in Figure \ref{scaling}. 

\section{Multiple Fields}
\label{multifield}

In this section we will generalize the bounds to the case of multiple fields. This is not as precise as we were able to make it in the context of single field slow-roll but the basic idea remains the same. This bound would apply to models of assisted inflation \cite{Liddle:1998jc} with $N$ scalar fields and the following action (note that in the following the sound speed of all the fields are set to 1).
\begin{equation}\label{generalactionforN}
S =  \int d^4 x \sqrt{g} \sum_i \left(\frac12\dot\phi_i^2\right) - W(\phi_i)
\end{equation}
In the simple case of no cross-terms between the fields, $W = \sum_i V(\phi_i)$. Such models were proposed in string theory \cite{Becker:2005sg, Dimopoulos:2005ac, Easther:2005zr} (see \cite{Kallosh:2007cc, Grimm:2007hs,  Misra:2007cq} for a recent discussion of their viability).  Assuming that there is a homogenous and isotropic solution of the fields' equations of motion, we can assume an FRW metric with Hubble scale given by
\begin{equation}
 \label{eq:Hgeneric}
3M_p^2H^2 = W(\phi_i) 
\end{equation}
Each field whose mass $m_{\phi_i}$ is smaller than $H$ will undergo quantum perturbations of order $\delta\phi_{i,q} \sim H$ where the subscript $q$ stands for quantum.  Neglecting perturbations of the metric, the action at second order is
\ba
S_2 &=& \int d^4x \sum_i \left(a^3 (\delta\dot\phi_{i,q})^2 + a (\p\delta\phi_{i,q})^2 + \right. \nonumber\\
&& \left. a^3 V_{,ii} (\delta\phi_{i,q})^2 + \cdots\right)
\ea
During inflation, the gradient term (and kinetic term) will be of order 
\ba
\sum_i H^4 \sim NH^4
\ea
since each field has perturbations of order $H$ over distance scales of order $1/H$. Demanding that this is smaller than the original background gives the following large $N$ bound
\ba
\frac{S_2}{S_0} \sim \frac{H^2 N}{M_p^2} < 1
\ea
Note that no conditions have been put on any of these scalars fields except that their masses are smaller that $H$ so that they fluctuate. So $N$ counts the number of scalar fields with low masses in the effective field theory, these fields do not need to have large vev or be``part" of the inflaton for this bound to hold. Finally note that this is a very simple derivation of a bound that has spurred many recent activities from black hole physics \cite{Dvali:2007hz, Dvali:2008sy}. A similar bound was derived in a different way in \cite{Ahmad:2008eu, Huang:2007st} while \cite{Huang:2007zt} used it to put constraint on eternal inflation in Large N models.

One can also look at higher order interactions in $S_3$. While we expect all sorts of couplings between 
the various fields, there will be, among others
\ba
S_3 = \int dt d^3x  a(t) \epsilon^{1/2} \sum_i \delta\phi_i \sum_j(\p\delta \phi_j)^2 + \dots
\ea
which gives
\begin{align}
\frac{S_3}{S_2} & \sim \epsilon^{1/2} \sum_i \delta\phi_i
\end{align}
But $\delta\phi_i$ is a random variable with mean zero and variance $H$. So the linear sum over $\delta\phi_i$ has a mean of zero and a variance of $\sqrt{N} H$. Using the variance to estimate the value of the contribution from $S_3$ gives the following condition:
\be
\frac{H\sqrt{N\epsilon}}{M_p} < 1
\ee

\section{Conclusions} 
\label{conclude}

In this paper, we have investigated the conditions under which the usual story of a slightly perturbed inflationary background, with the variance of the perturbations computed from the quadratic action in fluctuations, is valid. Requiring a perturbative calculation gives simple bounds on the Hubble scale, sound speed and the number of low mass fields. The bound on the Hubble scale is simply 
\ba
\frac{H^2}{M_p^2} < c_s^3
\ea
which for a small sound speed can require a significantly subplanckian Hubble scale.  A stronger bound on the sound speed can be found by looking at the higher order terms in the action. We found that
\ba
c_s^4 > \frac{H^2}{M_p^2c_s\epsilon} \sim \mathcal{P}_\zeta
\ea
When this bound is violated, perturbation theory breaks down since terms in $S_3$, $S_4$ and higher order all become of the same order as the quadratic terms. This is a signal that the $\zeta$ field is becoming a strongly interacting field whose variance cannot be computed by perturbative methods.  This bound also implies that a period of stochastic eternal inflation is always outside of the pertubative regime in models with sound speed less than 1 (really $1-c_s^2>\epsilon-2s$, for the terms in Eq. \ref{theS3term}).

When this bound on $c_s$ is translated into a bound for the magnitude of non-Gaussianity from the three-point function, we find that it is less stringent than the current experimental bound from CMB data. Nevertheless, for a scale dependent and decreasing sound speed, we find that perturbation theory may break down before one achieves the necessary number of e-folds for a successful inflationary era. We found that the number of e-folds before entering the non-perturbative regime is 
\be
N_e^{max}=\frac{1}{(4s-n_s+1)}\log\left[\frac{A(k_0)}{c_s^4(k_0)}\right]
\ee
Without an understanding of what happens in the non-perturbative regime, this can severely constrain models like DBI inflation where the sound speed varies. The scale dependent constraints (the bound on $c_s$ during the full range of inflation and some constraint on higher order interactions) must therefore be included in any analysis of this kind of model.

These constraints should be of interest in the popular program of reconstructing of the inflationary action. Very often, no constraints are imposed on the various terms in the action that is being reconstructed. We have shown here that there exist conditions that may be especially important for reconstructions involving a general kinetic term. It would be interesting further work to impose the consistency of perturbation theory in the general reconstruction approach in the literature, especially for models with general kinetic terms \cite{Bean:2008ga}. These constraints should also be imposed to restrict the phase space of possible configuration when studying such models with Monte-Carlo simulations \cite{Peiris:2007gz}.

We have been estimating the size of terms in the action to give these bounds. Of course, the formal definition of the breakdown of perturbation theory comes from looking at the contributions to correlation functions. Although we expect that the bounds we have found are similar to demanding that the 1-loop contribution to the two-point function is always small, it would be interesting to show this in detail for the general case. We also looked at a simple multi-field model of inflation and used the same logic to arrive at an upper bound on the number of fields.
\ba
N < \frac{M_p^2}{H^2}
\ea
This is exactly the same bound one finds from black hole physics or from renormalization of the Planck mass. Interestingly the black hole bound seems to indicate that the bounds we have found in this paper cannot be violated even in the strong coupling regime. Clearly, a more thorough analysis of the multi-fields case is needed and this is work in progress.

Finally, in this paper we have not discussed what happens when perturbation theory fails. Does inflation end, does the system dynamically correct, or is there an alternate description of the inflationary background? In the case where the higher order interactions are small $S_3<S_2$ but the gradient energy dominates over the background $S_2 > S_0$, we expect the universe to become inhomogeneous and stop inflating. It would be nice to have an explicit description of this transition. A promising way to answer these questions is to look at brane inflation where we can hope to use the string theory tools to get some handle on the non-perturbative regime.

\acknowledgments

We are particularly thankful to Richard Easther and Eugene Lim for early collaboration on this project. We also thank Puneet Batra, Marilena LoVerde, Liam McAllister, Alberto Nicolis, Andrew Tolley and Mark Wyman for valuable discussions. We thank Frederico Arroja, Kazuya Koyama and Andrei Linde for comments related to the first version of this paper. We would like to thank the organizers of the String and Cosmology workshop at KITPC where much progress was made on this work.  Finally we would like to thank the referee for many insightful comments that allowed us to improve the clarity and presentation of this paper. The work of L.L.~has been supported by the National Science Foundation under grant PHY-0505757 and the University of Texas A\&M. The work of S.S. is supported by the DOE under DE-FG02-92ER40699.

\providecommand{\href}[2]{#2}\begingroup\raggedright\endgroup

\end{document}